\documentclass[aps, prl, twocolumn]{revtex4-2}
\usepackage[utf8]{inputenc}
\usepackage[left=2.5cm,right=2.5cm]{geometry}\usepackage[utf8]{inputenc}
\usepackage{physics}
\usepackage{dsfont}
\usepackage{amsmath}
\usepackage{amssymb}
\usepackage{appendix}
\usepackage{braket}[]
\usepackage{xcolor}
\usepackage{bbold}
\usepackage{txfonts}
\usepackage{graphicx}

\begin{document}

\title{Breaking local quantum speed limits with steering}
\author{Federico Centrone}
\thanks{federico.centrone@icfo.eu.}
\affiliation{ICFO-Institut de Ciencies Fotoniques, The Barcelona Institute of Science and Technology, Mediterranean Technology Park, Avinguda Carl
Friedrich Gauss, 3, 08860 Castelldefels, Barcelona, Spain}
\author{Manuel Gessner}
\thanks{manuel.gessner@uv.es}
\affiliation{Departamento de F\'isica Te\'orica and IFIC, Universidad de Valencia-CSIC, C/ Dr Moliner 50, 46100 Burjassot (Valencia), Spain}
\begin{abstract}
We show how quantum correlations allow us to break the local speed limits of physical processes using only local measurements and classical communication between two parties that share an entangled state. Inequalities that bound the minimal time of evolution of a quantum state by energy fluctuations can be violated in the presence of steering by conditioning on the measurement outcomes of a remote system. Our results open up new pathways for studying how quantum correlations influence the dynamical properties of states and observables.
\end{abstract}

\maketitle

\textit{Introduction.--}Quantum theory unveils fundamental constraints and principles that impose limits on physical properties. The uncertainty principle, for instance, identifies limits on quantum fluctuations on non-commuting observables. Similarly, quantum speed limits identify bounds on the time scales of physical processes in terms of the available energy or its fluctuations. Apparent contradictions to such fundamental constraints often offer new insights into the nature of quantum systems. One such contradiction is exemplified by the EPR paradox \cite{einstein1935can}, which can be understood as a noncompliance of a local complementarity principle when properties are conditioned on the measurement outcomes of a distant system. 

Within the context of quantum information theory, we can attribute the EPR paradox to a strong form of quantum correlations, referred to as steering \cite{schrodinger1935discussion,wiseman2007steering, Cavalcanti_2016,Uola_2020}. These correlations, in addition to their fundamental significance, play a pivotal role in applications of quantum science, including one-sided device-independent quantum cryptography \cite{pawlowski2011semi} and subchannel discrimination \cite{piani2015necessary}. Experimentally implementable criteria to detect the EPR paradox and to reveal the presence of steering have been derived from uncertainty principles, entropic uncertainty relation and metrological complementarity \cite{cavalcanti2009experimental,costa2018steering, yadin2021metrological}.

In this Letter, we demonstrate that steering can lead to apparent violations of quantum speed limits. By conditioning the measurement results of a remote system, it becomes possible to achieve a quantum evolution with a characteristic time scale that exceeds the quantum speed limit of the local system. We derive steering criteria based on conditional speed limits derived from the Mandelstam and Tamm, as well as geometric approaches that leverage the quantum Fisher information as a metrological measure of quantum statistical speed. These results shed new light on the EPR paradox by linking it to the evolution time of quantum mechanical systems and identifying ways to speed up physical processes in post-selection using steering.

\textit{Quantum speed limits and the EPR paradox.--}The two main approaches towards deriving quantum speed limits are the \textit{minimal time approach} that relies on the effects of the statistical moments of the Hamiltonian that drives the system, and the \textit{geometric approach} that is based on the distinguishability of quantum states \cite{deffner2017quantum}. The first approach is based on an energy-time uncertainty relation by Mandelstam and Tamm \cite{mandelstam1991uncertainty}, which is derived from Heisenberg's uncertainty principle and the Ehrenfest theorem. Considering 
\begin{align}\label{eq:timescale}
    (\Delta \tau_{\hat M})_{\hat \rho}:= \frac{(\Delta \hat M)_{\hat \rho}}{|\frac{\partial }{\partial t}\langle \hat M \rangle_{\hat \rho}|}
\end{align}
the minimal time for the mean value of the observable $\hat M$ to change by a standard deviation yields the following minimal time quantum speed limit
  \begin{equation}\label{eq:mandelTamm}
    (\Delta \tau_{\hat M})_{\hat \rho}\geq \frac{\hbar}{2 (\Delta \hat H)_{\hat \rho}},
 \end{equation}
 which corresponds to the energy-time uncertainty relation. Here, $(\Delta \hat O)^2_{\hat \rho}=\langle \hat O^2\rangle_{\rho}-\langle \hat O\rangle^2_{\rho}$ is the variance of the observable $\hat O$ for the state $\hat\rho$, and $\hat H$ is the Hamiltonian generating the unitary evolution.

The EPR paradox was first introduced as an apparent incompatibility with a local uncertainty principle. Depending on the choice of the measurement basis in a remote system, the conditional quantum state is given either by a position eigenstate or a momentum eigenstate. Based on Heisenberg's uncertainty principle, a quantitative criterion for observation of an EPR paradox and  steering was first derived by Reid~\cite{reid1989demonstration}. 
Steering excludes the possibility of explaining the joint measurement results in terms of a local hidden state  (LHS) model.

A convenient description of the bipartite EPR steering scenario from Alice to Bob is given by assemblages $\mathcal{A}(a,\hat X)=p(a|\hat X)\hat\rho^{B}_{a|\hat X}$ that map a result $a$ for the measurement choice $\hat X$ by Alice to a non-normalized quantum state for Bob. Here, $p(a|\hat X)$ is the probability for Alice to obtain the result $a$ when measuring $\hat X$ and $\hat\rho^{B}_{a|\hat X}$ is the quantum state for Bob's subsystem conditioned on such an event. A local hidden state model exists if the measurement statistics can be modeled in terms of a set of local states $\hat \sigma^B_{\lambda}$, determined by the variable $\lambda$ that is distributed with probability $p(\lambda)$ and also determines the conditional probability $p(a|\hat X,\lambda)$, such that $\mathcal{A}(a,\hat X)=\sum_\lambda p(\lambda)p(a|\hat X,\lambda)\hat\sigma^B_\lambda$.

The conditional fluctuations for the observable $\hat O$ are given by $(\Delta \hat O)^2_{B|A}:=\min_{\hat X}\sum_a p(a|\hat X)(\Delta \hat O)^2_{\hat \rho^B_{a|\hat X}}$, i.e., by the average variance for each of the conditional states $\hat\rho^B_{a|\hat X}$ weighted by the probability of their occurrence. Alice's measurement setting is chosen such that the average fluctuations on Bob's side are minimized. Reid showed that conditional variances do not necessarily satisfy the Heisenberg uncertainty bound that is defined by the commutator of two local observables on Bob's system and any violation of this bound constitutes an EPR paradox and therefore indicates the presence of steering. Hence, even though the average variances necessarily obey the uncertainty principle, it is possible in the presence of steering, to sort and average the measured variances in post-selection using classical information about Alice's measurement choice and result. These conditional variances may break the uncertainty limit that is imposed by Heisenberg's uncertainty principle on Bob's local subsystem.

\begin{figure}
    \centering
    \includegraphics[width=\linewidth]{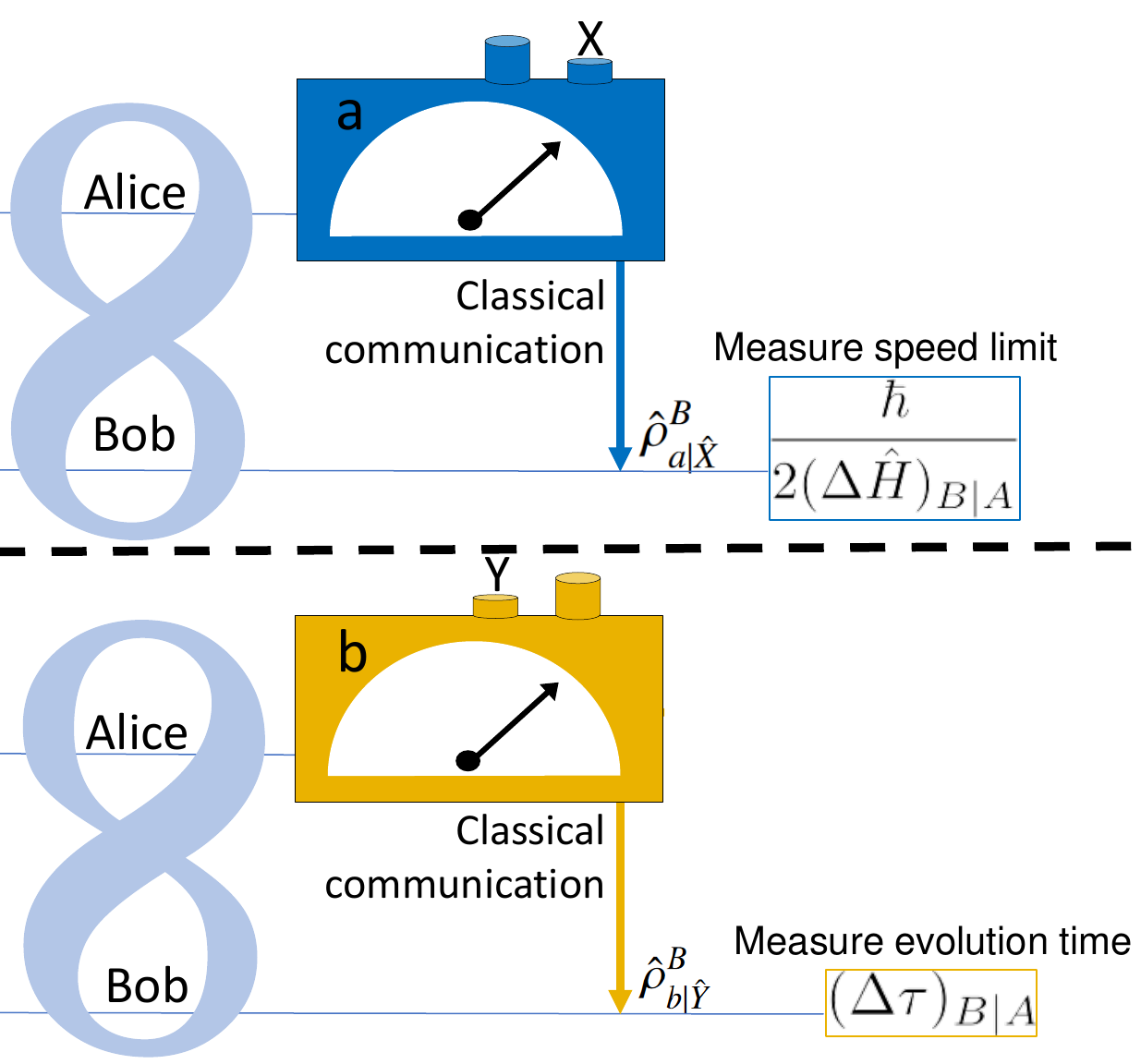}
    \caption{Conditioned on Alice’s measurement outcome and setting, Bob decides to either measure the QSL or the effective time of the evolution. The violation of the speed limit is a witness of quantum steering from Alice to Bob.}
    \label{fig:protocolScheme}
\end{figure}

\textit{
Breaking local quantum speed limit with steering.---}Let us first demonstrate that the EPR paradox can also manifest as a breach of a local quantum speed limit, assisted by steering and classical communication. To this end, we introduce the conditional characteristic time for variations of $\hat M$ as
\begin{align}\label{eq:condtime}
    (\Delta \tau_{\hat{M}})_{B|A}=\frac{(\Delta \hat M)_{B|A}}{|\frac{\partial }{\partial t}\langle \hat M \rangle|_{B|A}},
\end{align}
where $|\frac{\partial }{\partial t}\langle \hat M \rangle|_{B|A}=\max_{\hat X}\sum_a p(a|X) | \frac{\partial }{\partial t}\langle \hat M \rangle_{\hat\rho_{a|\hat X}^B}|$ are the conditional variations of the average value.
The conditional time scale~(\ref{eq:condtime}) is analogous to Eq.~(\ref{eq:timescale}) but both the fluctuations and the mean value variations are processed separately for each conditional state in post-selection.

Any assemblage that admits a LHS model must satisfy the local speed limit
\begin{align}\label{eq:SpeedReid}
    (\Delta \tau_{\hat{M}})_{B|A}\geq \frac{\hbar}{2(\Delta \hat H)_{B|A}},
\end{align}
for any local observable $\hat{M}$. The result is closely related to the Cavalcanti-Reid steering criterion~\cite{CavalcantiReid2007} that improves over Reid's original result~\cite{reid1989demonstration} by conditioning also on the result of the commutator~\cite{reid2009colloquium}; see the appendix for a proof. An interesting special case is found when $(\Delta \hat M)_{B|A}$ is time-independent. In the absence of steering, the time interval $\delta t=t_1-t_0$ that is needed to induce a change $\delta\langle\hat M\rangle_B:=|\langle\hat M\rangle_{\hat\rho^B(t_1)}-\langle\hat M \rangle_{\hat\rho^B(t_0)}|$ of the average value of $\hat M$ cannot be smaller than:
\begin{align}\label{eq:timeindSR}
    \delta t\geq \frac{\hbar}{2(\Delta\hat H)_{B|A}}\frac{\delta\langle \hat M\rangle_B}{(\Delta\hat M)_{B|A}}.
\end{align}
This bound can be further sharpened if the change in the expectation value of $\hat M$ is also conditioned on Alice's measurement results. A proof and additional details are provided in the appendix. For approaches that aim to link steering to alternative definitions of the quantum speed limit via the Wigner-Yanase skew information and different measures of coherence, see Ref.~\cite{mondal2017nonlocal}. In the following we focus on our result, that employs the most widely used notions of the quantum speed limit.

As an example, consider a two-mode continuous variable system that on Bob's side is subject to a translation generated by $\hat U(\delta t)=e^{-i\hat p \delta \langle\hat{x}\rangle_{B}/\hbar}$, where $\hat p$ is the momentum operator and an evolution time of $\delta t$ leads to a displacement of $\delta \langle\hat{x}\rangle_{B} =p_0\delta t/m$. If Alice and Bob share a two-mode squeezed state, upper bounds on the conditional variances of the two quadratures obtained from quadrature measurements are given by $ (\Delta \hat p_B)^2_{B|A} \leq (\Delta p_0)^2 z/(z^2 \sin^2(\theta)+\cos^2(\theta))$, $(\Delta  \hat x_B)^2_{B|A}\leq (\Delta x_0)^2 z/(z^2 \cos^2(\theta)+\sin^2(\theta))$, where the squeezing parameter $z$ ranges from $1$, implying no squeezing, to $0$, implying infinite squeezing. The parameter $\theta$ can be controlled via the transmissivity of a beam splitter that is used in the preparation of the state. Here, $(\Delta x_0)^2$ and $(\Delta p_0)^2$ are the variances of the quadratures of the state before applying the squeezing and they must satisfy the uncertainty principle $\Delta x_0\Delta p_0\geq\frac{\hbar}{2}$. We then consider the following protocol, depicted in Fig. \ref{fig:protocolScheme}: when Alice measures $\hat p$, Bob  also measures $\hat p$; when Alice measures $\hat x$, Bob displaces its part of the state for a time $\delta t$ and then measures $\hat x$. By repeating the experiment many times and post-selecting according to Alice's outcomes, Bob can infer the conditioned statistical moments of its quadratures. It's worth noting that for Gaussian states, quadrature measurements are not always the optimal choice, as indicated in \cite{benech2022einstein}. Nevertheless, they remain the most practical option for implementation in quantum optics and are well-suited for description within the framework of symplectic geometry. In this case, it is sufficient to apply Eq.~(\ref{eq:timeindSR}) to obtain
\begin{equation}
    \delta t\geq\frac{ \hbar m }{2p_0 (\Delta\hat p)_{B|A}}\frac{\delta \langle\hat x\rangle_{B}}{(\Delta \hat x)_{B|A}}.
\end{equation}
It can be easily seen that if $\theta=0$, the time of the evolution is simply given by the ratio of the mean position with the velocity. Conversely, for $\theta>0$ and $z<1$ the state is entangled, and the bound is always violated.

As another case study, consider that of a free-propagating entangled particle with $\hat H=\hat p^2/2m$. Assuming that the initial average position of the particle is $\langle \hat x_B(0)\rangle=0$ and the mean momentum (conserved through the evolution) is $p_0$, we can derive a local bound for the evolution of an LHS model (for an explicit expression see the appendix). In Fig.~\ref{fig:freeP} we show when the condition 
\begin{align}\label{eq:gamma}
\gamma:=\frac{2(\Delta\tau_{\hat{M}})_{B|A}(\Delta\hat H)_{B|A}}{\hbar}\geq 1.
\end{align}
is violated, implying steering, for the free particle with $\hat M=\hat x_B$. In addition, computing the evolution in time of the variance of $\hat x_B$, we can isolate $\delta t$ and find the following bound for the physical time of the local evolution (see proof in appendix):
\begin{equation}
    \delta t\geq \frac{(1-\gamma)\hbar}{4R^2(\Delta\hat H)^2_{B|A}},
\end{equation}
where any violation again implies steering, and $R=\Delta p_0/p_0$. This criterion becomes increasingly harder to violate as time evolves due to the spreading of the wavefunction under free evolution. The tightest bound is obtained at $\delta t =0$ and in this case it is equivalent to the one obtained from Eq.~(\ref{eq:gamma}).
\begin{figure}
    \centering
    \includegraphics[width=\linewidth]{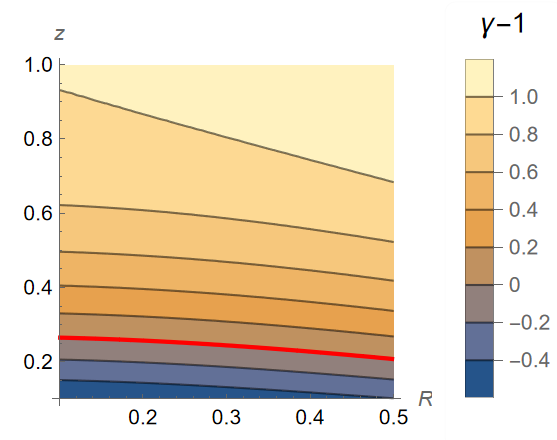}
    \caption{Free particle violating local speed limit with steering. Density plot of the condition $\gamma-1>0$ as a function of squeezing $z$ and the ratio $R=\Delta p_0/p_0$ between the uncertainty and the mean value of the momentum of the initial state, for $\theta=\pi/4$.}
    \label{fig:freeP}
\end{figure}

\textit{Geometric speed limits.---}Geometric approaches to quantum speed limits measure the  the rate of change of a quantum state or an observable at any given instance of a trajectory between two points. By identifying optimal trajectories and determining the fastest possible approach from one point in state space to another, speed limits can be determined~\cite{taddei2013quantum, deffner2017quantum}. These approaches are based on a geometric notion of distance between states in Hilbert space, such as the Bures angular distance 
$
D[\hat \rho,\hat  \sigma]:=\arccos(\Tr[\sqrt{\sqrt{\hat\rho}\hat\sigma\sqrt{\hat\rho}}])$ \cite{bengtsson2017geometry}. For infinitesimal system evolution $\hat \rho(t)\rightarrow\hat  \rho(t+\epsilon)$, the rate of change in state space is given by $
v(t):=\lim_{\epsilon\rightarrow 0} \frac{D(\hat \rho(t+\epsilon),\hat \rho(t))}{\epsilon}=\frac{1}{2\hbar}\sqrt{F_Q[\hat \rho(t)]} $ \cite{taddei2013quantum,deffner2017quantum}. Here, $F_Q[\hat \rho(t)]$ is the quantum Fisher information of the state $\hat \rho(t)$ with respect to changes of the time parameter $t$. This quantity further determines the ultimate precision limit for parameter estimation protocols that aim to extract the parameter $t$ through measurements on $\hat{\rho}(t)$~\cite{braunstein1994statistical}. 

In the scenario presented in~\cite{yadin2021metrological}, a steering-assisted phase-estimation protocol is employed, where Alice communicates the measurement setting $\hat X$ and outcome $a$ to Bob. Bob can adapt the observable $\hat X$ based on the conditional state $\hat \rho^B_{a|\hat X}$ to maximize sensitivity. After multiple rounds of the protocol, Bob can achieve an average sensitivity characterized by the conditional quantum Fisher information $F_{B|A}[\hat{H}]:=\max_{\hat X} \sum_a p(a|\hat X)F_Q\left[\hat \rho^B_{a|\hat X},\hat{H}\right]$. In the case of a unitary evolution generated by $\hat H$ the quantity $F_Q\left[\hat \rho(t)\right]=F_Q\left[\hat \rho^B_{a|\hat X},\hat{H}\right]$ becomes independent of $t$. It can be shown that for any assemblage $\mathcal{A}$ admitting a LHS model, the following bound holds: $F_{B|A}[\hat{H}]\leq 4(\Delta \hat H)^2_{B|A}$.

Defining the conditional mean quadratic speed as $\langle v^2 \rangle_{B|A}:= \max_{\hat X}\sum_a p(a|\hat{X})\frac{F_Q\left[\hat \rho^B_{a|\hat{X}},\hat{H}\right]}{4\hbar^2 }$ the conditional rate of change in the local state space of an assemblage described by an LHS model is bounded by~\cite{yadin2021metrological}
\begin{equation}\label{eq:speedBound}
\hbar^2 \langle v^2 \rangle_{{B|A}}\leq (\Delta\hat H)^2_{B|A}.
\end{equation}
This bound remains valid when integrated over time, allowing for a time average on both sides of the equation. For the unitary dynamics we are considering, this leads to the following final bound:
\begin{equation}\label{eq:metroTime}
\delta t \geq \hbar \sqrt{\frac{\langle D(\delta t)^2 \rangle _{B|A} }{(\Delta\hat H)^2_{B|A}}}.
\end{equation}
where $\delta t$ is the physical time Bob's conditioned system is left to evolve before being measured and $\langle D(\delta t)^2 \rangle _{B|A}=\max_{\hat X}\sum_a p(a|\hat{X})D\left(\hat\rho^B_{a|\hat X}(\delta t), \hat\rho^B_{a|\hat X}(0)\right)^2$ is the average quadratic distance between the initial and final state on Bob's side after conditioning on Alice's outcome. The bound above  can be violated in the presence of steering and reduces to the standard QSL for arbitrary states and unitary evolution \cite{uhlmann1992energy,pfeifer1993fast}. This type of bound was recently showed to set a fundamental bound on the minimum attainable phase estimation error through the quantum Cramér–Rao bound, relating the precision directly to the underlying dynamics of the system \cite{maleki2023speed}. The proof of the above relation can be found in the appendix.

Let us now consider a multipartite quantum state $\hat \rho$ with discrete degrees of freedom prepared in the form of a GHZ state with limited visibility $p$. In this case, Alice controls one qubit of the state and will measure it either in the $\hat \sigma_x$ or the $\hat \sigma_z$ basis. According to Alice's measurement setting, Bob will either let the system evolve for a time $\delta t$ with the Hamiltonian $\hat J_B^z = \frac{\mu}{2}\sum_i \hat\sigma_z^{(i)}$ or measure the energy variance of the state. By employing equation \ref{eq:metroTime}, we can derive a steering condition on the time of evolution, given by:

\begin{equation}\label{eq:GHZtime}
\delta t\geq \frac{\hbar}{\mu} p \sqrt{\frac{N }{\left(2^{1-N}(1-p)+p\right)\left(1-p+N(1-p)p\right)}}.
\end{equation}

This is a bound on the physical time of evolution of the quantum state. It corresponds to a time-integrated formulation of the geometric criterion provided in~\cite{yadin2021metrological}. This new formulation opens up a path towards witnessing quantum steering based on the time evolution of a system that could be useful for some experimental platforms~\cite{del2021probing, pires2023experimental}.

\textit{Conclusions---} In this work, we have established a unified framework that links steering and quantum speed limits, opening up new applications for investigating quantum correlations. By introducing steering witnesses based on mechanical quantities, such as the time of flight of an entangled particle in rectilinear uniform motion, we have introduced novel methods to certify quantum correlations in mechanical and thermodynamical scenarios. Our findings also suggest that steering can potentially provide a quantum advantage in mechanical tasks, such as accelerating certain physical processes in post-selection.

A promising direction for further exploration would involve applying our framework to the ballistic scenario described in \cite{trillo2022quantum}, to investigate the effect of steering on entangled projectiles and quantum backflow. Additionally, it would be intriguing to study the impact of nonclassical correlations in other scenarios where quantum speed limits are commonly employed, such as optimal quantum control systems \cite{caneva2009optimal}, the entropy production rate of quantum engines \cite{deffner2010generalized}, the maximal power of quantum battery arrays \cite{campaioli2017enhancing}, the Landauer principle in finite time \cite{zhen2021universal}, or the minimum time of a quantum measurement \cite{shettell2022bounding}. 

\begin{acknowledgments}
\textit{Acknowledgments.---} We thank Matteo Scandi for fruitful discussions. This work was funded by MCIN/AEI/10.13039/501100011033 and the European Union 'NextGenerationEU' PRTR fund [RYC2021-031094-I]. This work has been founded by the Ministry of Economic Affairs and Digital Transformation of the Spanish Government through the QUANTUM ENIA project call—QUANTUM SPAIN project, by the European Union through the Recovery, Transformation and Resilience Plan—NextGenerationEU within the framework of the Digital Spain 2026 Agenda, and by the CSIC Interdisciplinary Thematic Platform (PTI+) on Quantum Technologies (PTI-QTEP+). This work has been funded by the Government of Spain (Severo Ochoa CEX2019-000910-S and European Union NextGenerationEU PRTR-C17.I1) and European Union (PASQuanS2.1, 101113690), Fundació Cellex, Fundació Mir-Puig, Generalitat de Catalunya (CERCA).
\end{acknowledgments}

       \bibliographystyle{ieeetr}
\bibliography{biblio}

\vspace{2cm}

\textbf{Appendix/Supplementary.}
\textit{Proof of Eq.~(\ref{eq:SpeedReid})}. In the presence of a LHS model, we use $\hat\rho^B_{a|\hat X}=\sum_{\lambda}p(\lambda)p(a|\hat X,\lambda)\hat\sigma^B_{\lambda}/p(a|\hat X)$, the concavity of the variance, and $\sum_a p(a|\hat X,\lambda)=1$, to obtain
\begin{align}
    (\Delta \hat M)_{B|A}^2\geq \sum_{\lambda} p(\lambda)(\Delta \hat M)^2_{\hat \sigma^B_{\lambda}}.
\end{align}
From the Cauchy-Schwartz inequality and the local speed limit~(\ref{eq:mandelTamm}) for the state $\hat\sigma^B_{\lambda}$, we then find
\begin{align}
    (\Delta \hat M)_{B|A}^2(\Delta \hat H)_{B|A}^2\geq \left|\sum_{\lambda}p(\lambda)\frac{\hbar}{2}\left|\frac{\partial}{\partial t}\langle \hat M\rangle_{\hat \sigma^B_{\lambda}}\right|\right|^2.
\end{align}
Moreover, 
\begin{align}\label{eq:eq2}
\sum_a  \left| \frac{\partial }{\partial t}\sum_{\lambda}p(\lambda)p(a|\hat X,\lambda)\langle \hat M \rangle_{\hat\sigma^B_{\lambda}}\right|\leq \sum_{\lambda}p(\lambda)\left|\frac{\partial }{\partial t}\langle \hat M \rangle_{\hat\sigma^B_{\lambda}}\right|.
\end{align}
Since this bound holds for any choice of $\hat X$, this includes the choice that maximizes the conditional variations of the average, and therefore
\begin{align}\label{eq:eq1}
    \left|\frac{\partial }{\partial t}\langle \hat M \rangle\right|_{B|A}\leq\sum_{\lambda}p(\lambda)\left|\frac{\partial }{\partial t}\langle \hat M \rangle_{\hat \sigma^B_{\lambda}}\right|
\end{align}
Inserting Eq.~(\ref{eq:eq1}) into Eq.~(\ref{eq:eq2}), we thus obtain the bound~(\ref{eq:SpeedReid}) for LHS models.

\textit{Proof of Eq.~(\ref{eq:timeindSR}).} From the triangle inequality, we obtain
\begin{align}
    &\qquad\int_{t_0}^{t_1} dt|\frac{\partial }{\partial t}\langle \hat M \rangle|_{B|A}\notag\\&=\int_{t_0}^{t_1} dt\max_{\hat X}\sum_a p(a|\hat X) \left| \frac{\partial }{\partial t}\langle \hat M \rangle_{\hat\rho_{a|\hat X}^B}\right|\notag\\
    &=\max_{\hat X}\sum_a p(a|\hat X) \int_{t_0}^{t_1} dt\left| \frac{\partial }{\partial t}\langle \hat M \rangle_{\hat\rho_{a|\hat X}^B}\right| \notag\\
    &\geq \delta\langle \hat M\rangle_{B|A},
\end{align}
where we have defined
\begin{align}
    \delta\langle \hat M\rangle_{B|A}=\max_{\hat X}\sum_a p(a|\hat X) \left| \langle\hat M \rangle _{\hat\rho_{a|\hat X}^B(t_1)} -\langle\hat M \rangle _{\hat\rho_{a|\hat X}^B(t_0)} \right|
\end{align}
Applying one more time the triangle inequality to bring the summation inside the absolute value, we obtain a looser but simpler bound
\begin{equation}
    \delta\langle \hat M\rangle_{B|A}\geq \delta\langle\hat M\rangle_B=|\langle\hat M\rangle_{\hat\rho^B(t_1)}-\langle\hat M \rangle_{\hat\rho^B(t_0)}|.
\end{equation}
This final bound only depends on Bob's reduced state and is independent of the assemblage.

Note that the fluctuations of $\hat H$ are always time-independent for unitary evolution. If $(\Delta \hat M)_{B|A}$ is also independent of time, we can integrate Eq.~(\ref{eq:SpeedReid}) to obtain the result Eq.~(\ref{eq:timeindSR}).

\textit{Speed steering with a free particle}. \label{appendix2} 
Since both the state and the dynamics are Gaussian, we can employ the formalism of symplectic geometry. The initial covariance matrix of the two modes squeezed state shared between Alice and Bob is
\begin{equation}\label{eq:cov}   V_{AB}=\begin{pmatrix} \sigma_A &  \sigma_{AB}\\ \sigma_{AB} & \sigma_B
   \end{pmatrix}
\end{equation}
with
{\small
\begin{equation}
    \sigma_A=\begin{pmatrix}
        (\Delta x_0)^ 2\left(z \cos^{2}{\left(\theta \right)} + \frac{\sin^{2}{\left(\theta \right)}}{z}\right) & 0 \\ 0 & (\Delta p_0)^2\left( z \sin^{2}{\left(\theta \right)} + \frac{\cos^{2}{\left(\theta \right)}}{z}\right)
    \end{pmatrix},\notag
\end{equation}
\begin{equation}
    \sigma_B=\begin{pmatrix}
        (\Delta x_0)^ 2\left(z \sin^{2}{\left(\theta \right)} + \frac{\cos^{2}{\left(\theta \right)}}{z}\right)  & 0 \\ 0 & (\Delta p_0)^2\left( z \cos^{2}{\left(\theta \right)} + \frac{\sin^{2}{\left(\theta \right)}}{z}\right)
    \end{pmatrix},\notag
\end{equation}
\begin{equation}\label{eq:covAB}
    \sigma_{AB}=\begin{pmatrix}  (\Delta x_0)^ 2\frac{\hbar^2}{p_0^2}\left(\sin{\left(2\theta \right)}\left(
          \frac{1}{2z} -\frac{z}{2}\right)\right) & 0 \\ 0 &  (\Delta p_0)^2\left( \sin{\left(2\theta \right)}\left(
       \frac{z}{2}   - \frac{1}{2z}\right)\right)
    \end{pmatrix},
\end{equation}
}
where  $(\Delta x_0)^ 2$ and $(\Delta p_0)^2$ are the variances of quadratures of the state with no squeezing. To satisfy the uncertainty principle they yield $\Delta x_0\Delta p_0=\frac{\hbar}{2}(k+1)$, for some non-negative $k$ that expresses thermal fluctuations beyond the vacuum. Note that for a mixing angle of $\theta=\pi/4$ we recover the usual form of the two-mode squeezed vacuum state with squeezing parameter $r$ upon substituting $z=e^{-2r}$. Bob's state after Alice's measurement is described by the first two moments of its quadratures  $\vec {\hat{q}}_B=(\hat x_B,\hat p_B)^T$, which need to be updated as \cite{serafini2017quantum}
\begin{equation}\label{eq:condMoments}
    \begin{cases}
        \langle \vec{ \hat{ q}}_B \rangle _{\hat{\rho}^B_{\vec{a}|\sigma_{\hat M}}}&=  \langle \vec{ \hat{ q}}_B \rangle_{\hat \rho^B}+\sigma_{AB}(\sigma_A+\sigma_{\hat M})^+(\vec{a} - \langle \vec{ \hat{ q}}_A \rangle_{\hat\rho^A}) \\
        (\sigma_{B})_{\vec{a}|\sigma_{\hat M}}&=\sigma_B-\sigma_{AB}(\sigma_A+\sigma_{\hat M})^+\sigma_{AB}^T
    \end{cases},
\end{equation}
where $\vec a $ and $\sigma_{\hat M}$  are Alice's measurement outcome and the post-measurement covariance matrix for the setting $\hat M$ respectively, while $O^+$ denotes the pseudoinverse (or the Moore-Penrose inverse) of the matrix $O$. In particular, $\sigma_{\hat M}$ completely specifies the type of measurement and the noise model in the system and its  form is independent of the outcome $\vec a$. For ideal homodyne measurements of Alice's quadratures $\hat x_A$ or $\hat p_A$ we have $\sigma_{\hat x_A}=\underset{z\rightarrow 0}{\lim}\;\text{diag}\left( z, 1/z\right)$ and  $\sigma_{\hat p_A}=\underset{z\rightarrow \infty}{\lim}\;\text{diag}\left( z, 1/z\right)$. This yields
\begin{align}
    (\sigma_A+\sigma_{\hat x_A})^+&=\underset{z\rightarrow 0}{\lim}\;\left(\sigma_A+\text{diag}\left( z, 1/z\right)\right)^+=
    \begin{pmatrix}
       1/ \sigma_A^{(1,1)} & 0\\ 0 & 0
    \end{pmatrix},\notag\\ (\sigma_A+\sigma_{\hat p_A})^+&=\underset{z\rightarrow \infty}{\lim}\;\left(\sigma_A+\text{diag}\left( z, 1/z\right)\right)^+=
    \begin{pmatrix}
        0 & 0\\ 0 & 1/\sigma_A^{(2,2)}
    \end{pmatrix},
\end{align}
while 
 $\det(\sigma_A+\sigma_{\hat x_A})= \sigma_A^{(1,1)}$ for an $\hat x_A$ measurement and  $\det(\sigma_A+\sigma_{\hat{p}_A})= \sigma_A^{(2,2)}$ for a $\hat p_A$ measurement. 

The probability for Alice to obtain the result $\vec a$ is given by
\begin{equation}\label{eq:outcomeProbability}
    p(\vec a|\sigma_{\hat M})=\frac{e^{-( \vec{a} - \langle \vec{ \hat{ q}}_A \rangle_{\hat{\rho}^A})^T (\sigma_A+\sigma_{\hat M})^+( \vec{a} - \langle \vec{ \hat{ q}}_A \rangle_{\hat{\rho}^A})/2}}{\sqrt{2\pi\det(\sigma_A+\sigma_{\hat M})}},
\end{equation}
from which we can compute bounds on the conditional speed, variance of the energy, and variance of the position from homodyne measurements as:
\begin{align}
    \left|\frac{\partial}{\partial t} \langle  \hat{ x}_B \rangle\right|_{B|A}&\geq\int d\vec{a}p(\vec a|{\sigma_{\hat M}})\left|\frac{\partial}{\partial t} \langle  \hat{ x}_B \rangle_{\hat{\rho}^B_{\vec{a}|\sigma_{\hat M}}}\right|,\\
     (\Delta \hat x_B)^2_{B|A}&\leq\int d\vec{a}p(\vec a|\sigma_{\hat M})(\Delta \hat x_B)^2_{\hat\rho^B_{\vec{a}|\sigma_{\hat M}}},\\
    (\Delta\hat H)^2_{B|A}&\leq\int d\vec{a}p(\vec a|{\sigma_{\hat M}})(\Delta\hat H)^2_{\hat \rho^B_{\vec{a}|\sigma_{\hat M}}}.
\end{align}
We remember here that Alice's measurement has not been systematically optimized. However, the quadrature measurements considered here describe a natural physical scenario and, as will be shown in the following, suffice to demonstrate steering.

Using the Ehrenfest relation, $\frac{\partial}{\partial t} \langle  \hat{ x}_B \rangle = \langle  \hat{ p}_B \rangle/m$, and the triangle inequality we can bound the conditional speed by
\begin{align}
    \left|\frac{\partial}{\partial t} \langle  \hat{ x}_B \rangle\right|_{B|A}&\geq\frac{1}{m}\int d\vec{a}p(\vec a|{\sigma_{\hat M}})\left|\langle  \hat{p}_B \rangle_{\hat{\rho}^B_{\vec{a}|\sigma_{\hat M}}}\right|\notag\\&\geq\frac{1}{m}\left|\int d\vec{a}p(\vec a|{\sigma_{\hat M}})\langle  \hat{p}_B \rangle_{\hat{\rho}^B_{\vec{a}|\sigma_{\hat M}}}\right|.
\end{align}
Given that the covariance matrix~(\ref{eq:covAB}) does not contain any correlations between $\hat x$ and $\hat p$, when Alice measures $\hat x$, according to Eq.~(\ref{eq:condMoments}), it will not affect the value of $\hat p$ on Bob's side, hence
$\langle \hat{p}_B \rangle_{\hat{\rho}^B_{\vec{a}|\hat \sigma_{\hat{x}}}}=\langle \hat{p}_B \rangle_{\hat{\rho}^B}=:p_0$ is constant and yields Bob's average momentum of the initial state. We thus obtain
\begin{equation}\label{eq:condXderivative}
    \left|\frac{\partial}{\partial t} \langle  \hat{ x}_B \rangle\right|_{B|A}\geq\frac{|p_0|}{m}.
\end{equation}


We use Eq.~(\ref{eq:condMoments}) to obtain the following upper bounds on the conditional variances from homodyne measurements of $\hat{x}_A$ and $\hat{p}_A$, respectively:
\begin{align}
    (\Delta \hat{x}_B)^2_{B|A}&\leq \sigma_B^{(1,1)}-\frac{(\sigma_{AB}^{(1,1)})^2}{\sigma_A^{(1,1)}},\\
    (\Delta \hat{p}_B)^2_{B|A} &\leq \sigma_B^{(2,2)}-\frac{(\sigma_{AB}^{(2,2)})^2}{\sigma_A^{(2,2)}}.
\end{align}
For the covariance matrix~(\ref{eq:cov}), we obtain
\begin{align}
(\Delta \hat{x}_B)^2_{B|A}&\leq(\Delta x_0)^ 2\frac{z}{z^2 \cos^2(\theta)+\sin^2(\theta)},\\
    (\Delta \hat{p}_B)^2_{B|A} &\leq(\Delta p_0)^2 \frac{z}{z^2 \sin^2(\theta)+\cos^2(\theta)}.
\end{align}

The characteristic time scale for variations of $\hat{x}_B$ at the initial moment of the evolution is therefore bounded from above by
\begin{align}\label{eq:tauxB}
    (\Delta\tau_{\hat{x}_B})^2=\frac{(\Delta \hat{x}_B)^2_{B|A}}{\left|\frac{\partial}{\partial t} \langle  \hat{ x}_B \rangle\right|^2_{B|A}}&\leq \frac{m^2}{p_0^2}\left(\sigma_B^{(1,1)}-\frac{(\sigma_{AB}^{(1,1)})^2}{\sigma_A^{(1,1)}}\right)\notag\\
    &=\frac{m^2 (\Delta x_0)^2}{p_0^2} \frac{z}{z^2 \cos^2(\theta)+\sin^2(\theta)}.
\end{align}

To determine the local speed limit~(\ref{eq:SpeedReid}), i.e., an upper bound for this time scale that holds for all LHS models, we need to obtain the conditional variance of the Hamiltonian $(\Delta\hat H)^2_{B|A}=\frac{1}{4m^2}(\Delta \hat p_B^2)^2_{B|A}$. Here,
\begin{align}\label{eq:intcond4thmom}
    (\Delta \hat p_B^2)^2_{B|A}&\leq \int d\vec{a}p(\vec{a}|\sigma_{\hat{M}})(\Delta\hat{p}_B^2)^2_{\hat{\rho}^B_{\vec{a}|\sigma_{\hat{M}}}}
\end{align}
is again an upper bounded from a homodyne setting specified by $\sigma_{\hat{M}}$. To compute $(\Delta\hat{p}_B^2)^2_{\rho^B_{\vec{a}|\sigma_{\hat{M}}}}$, we exploit the properties of Gaussian integrals, namely the moment generating function of a Gaussian distribution. The fourth moment of $\hat{p}_B$ reads:
\begin{align}
    (\Delta\hat p_B^2)^2_{\hat{\rho}^B_{\vec{a}|\hat M}}&=\langle \hat p_B^4\rangle_{\hat{\rho}^B_{\vec{a}|\hat M}} -\langle \hat p_B^2\rangle_{\hat{\rho}^B_{\vec{a}|\hat M}}^2\notag\\&=4\langle \hat p_B\rangle^2_{\hat{\rho}^B_{\vec{a}|\hat M}}(\Delta \hat p_B)^2_{\hat{\rho}^B_{\vec{a}|\hat M}}+2(\Delta \hat p_B)^4_{\hat{\rho}^B_{\vec{a}|\hat M}}.
\end{align}
Let us now consider a momentum measurement by Alice, i.e., $\hat{M}=\hat{p}_A$. Using
\begin{align}
    p(a|\sigma_{\hat p_A})=\frac{e^{\frac{-(a-\langle \hat p_A\rangle_{\hat{\rho}^A})^2}{2\sigma_A^{(2,2)}}}}{\sqrt{2\pi \sigma_A^{(2,2)}}},
\end{align}
and Eq.~(\ref{eq:condMoments}), the integral on the right-hand side of~(\ref{eq:intcond4thmom}) yields
\begin{align}
    &\quad\int dap(a|\sigma_{\hat{p}_A})(\Delta\hat{p}_B^2)^2_{\hat{\rho}^B_{a|\sigma_{\hat{p}_A}}}\\
    &=2\left(\sigma_B^{(2,2)}-\frac{(\sigma_{AB}^{(2,2)})^2}{\sigma_A^{(2,2)}}\right)\left(\frac{(\sigma_{AB}^{(2,2)})^2}{\sigma_A^{(2,2)}}+2p_0^2+\sigma_B^{(2,2)}\right).\notag
\end{align}
\begin{widetext}
For the covariance matrix~(\ref{eq:cov}), we obtain
\begin{align}
    \int dap(a|\sigma_{\hat{p}_A})(\Delta\hat{p}_B^2)^2_{\hat{\rho}^B_{a|\sigma_{\hat{p}_A}}}=
    \frac{ { \Delta  p_0 }^2 \left(\left(z^2+1\right) \left(4  { p_0}^2 z+ { \Delta  p_0 }^2 \left(z^2+1\right)\right)-4  { p_0}^2 z \left(z^2-1\right) \cos (2 \theta )+ { \Delta  p_0 }^2 \left(z^2-1\right)^2 \cos (4 \theta )\right)}{2 \left(\cos ^2(\theta )+z^2 \sin ^2(\theta )\right)^2}
\end{align}
and the speed limit~(\ref{eq:SpeedReid}) is bounded from below by
\begin{align}\label{eq:csl}
     \frac{\hbar^2}{4(\Delta\hat H)^2_{B|A}}\geq \frac{m^2 \hbar ^2 \left(\cos ^2(\theta )+z^2 \sin ^2(\theta )\right)^2}{2 \text{$\Delta $p}_0^2 \left(\left(z^2+1\right) \left(4 p_0^2 z+\text{$\Delta $p}_0^2 \left(z^2+1\right)\right)-4 p_0^2 z \left(z^2-1\right) \cos (2 \theta )+\text{$\Delta $p}_0^2 \left(z^2-1\right)^2 \cos (4 \theta )\right)}.
\end{align}
Using Eqs.~(\ref{eq:tauxB}) and~(\ref{eq:csl}), we find that a sufficient condition for the violation of the local speed limit $(\Delta\tau_{\hat{x}_B})^2-\frac{\hbar^2}{4(\Delta\hat H)^2_{B|A}}\geq 0$ is given when $\gamma-1\geq 0$, where $\gamma:=\frac{(\Delta\tau_{\hat{x}_B})2(\Delta\hat H)_{B|A}}{\hbar}$. For the covariance matrix~(\ref{eq:cov}) we have 
\begin{align}\label{eq:gammaEXP}
    \gamma^2=\frac{(k+1)^2 z \left(-R^2 \left(1-z^2\right)^2 \cos (4 \theta )+\left(z^2+1\right) \left(R^2 \left(z^2+1\right)+4 z\right)+4 z \left(1-z^2\right) \cos (2 \theta )\right)}{2 \left(\sin ^2(\theta )+z^2 \cos ^2(\theta )\right) \left(\cos ^2(\theta )+z^2 \sin ^2(\theta )\right)^2},
\end{align}
where we used $\Delta x_0\Delta p_0=\frac{\hbar}{2}(k+1)$ and $R=\frac{\Delta p_0}{p_0}$. When the above quantity becomes smaller than $1$ we have steering. Interestingly, this quantity only depends on the ratio between the variance and the mean value of the momentum of the initial state of the particle. We have for the separable case $\theta=0$
\begin{equation}
   \gamma^2=2 (k+1)^2 \left(R^2 z+2\right),
\end{equation}
which admits a LHS for all physical values of the parameters. For the entangled case $\theta=\pi/4$ we have
\begin{equation}
  \gamma^2=\frac{8 (k+1)^2 z \left(R^2 \left(z^4+1\right)+2 \left(z^3+z\right)\right)}{\left(z^2+1\right)^3}.
\end{equation}
In this case, we have an interplay between the various parameters. In particular, $\gamma$ decreases with the purity of the initial state, with larger squeezing and for a small value of the ratio between the precision and the mean value of the momentum. This is plotted in \ref{fig:freeP}.
\end{widetext}

\textit{Proof of the physical time bound for the free particle.}\label{appendix:physicaltime}
If Alice measures the quadrature $\hat x$, Bob will let his particle propagate freely with the quadratic Hamiltonian $\hat H=\frac{1}{2}\vec {\hat{q}}^T h \vec {\hat{q}}=\hat p^2/2m$, where $h=\begin{pmatrix}
    0 & 0\\ 0 & 1/m
\end{pmatrix}$. 
The evolution of the first two moments of the quadratures after a time $\delta t$ is implemented by the symplectic transformation $S_{\delta t}=e^{\Omega h \delta t}$, where the symplectic form reads $\Omega=\begin{pmatrix}
    0 & 1 \\ -1 & 0
\end{pmatrix}$. 
We have then
{\small
\begin{equation}
    \begin{cases}
        \langle \vec{\hat q}(t)-\vec{\hat q}(0)\rangle_{B|A} = S_{\delta t}\langle \vec{\hat q}(0)\rangle_{B|A}-\langle \vec{\hat q}(0)\rangle_{B|A} = (p_0\delta t/m, p_0)^T\\
        \sigma(t)_{B|A}=S_{\delta t}\sigma(0)_{B|A}S_{\delta t}^T=\begin{pmatrix}
            (\Delta \hat x)^2_{B|A}+(\Delta\hat p_B)^2\delta t^2/m^2 & (\Delta \hat p_B)^2\delta t /m \\ (\Delta\hat p_B)^2\delta t/m & (\Delta\hat p_B)^2
        \end{pmatrix}
    \end{cases},
\end{equation}
}
from which we get the variance of the position evolved in time
\begin{equation}\label{eq:condVarTime}
    (\Delta \hat x_{B|A})^2(\delta t) =(\Delta \hat x)^2_{B|A}(0)+(\Delta\hat p_B)^2(\delta t/m)^2 .
\end{equation}
By plugging this into definition~(\ref{eq:condtime}), the condition $\frac{(\Delta\tau_{\hat{x}_B})^2(t)4(\Delta\hat H)^2_{B|A}}{\hbar^2}>1$ will become
\begin{equation}
    \delta t\geq \frac{(1-\gamma)\hbar}{4R^2(\Delta\hat H)^2_{B|A}},
\end{equation}
where $\gamma$ and $R$ were defined in the previous appendix.

\textit{Proof of Eq.~(\ref{eq:metroTime}).}
We bound the integral
    $\int^{\delta t}_0 dt \langle v (t)^2\rangle $,
from below, where 
\begin{equation}
    v(t)= \lim_{\epsilon \rightarrow 0} \frac{D(\hat \rho(t+\epsilon),\hat \rho(t))}{\epsilon}.
\end{equation}
Let us consider the Cauchy-Schwartz inequality
\begin{equation}
    \int_0^{\delta t} |f(t)|^2dt\int_0^{\delta t} |g(t)|^2dt\geq \left| \int_0^{\delta t} f(t)g^*(t) dt\right|^2,
\end{equation}
with $f(t)=v(t)$ and $g(t)=\frac{1}{\sqrt{\delta t}}$, to obtain
\begin{equation}
     \int_0^{\delta t} v(t)^2dt\geq \frac{1}{\delta t}\left|\int_0^{\delta t} v(t) dt \right|^2=\frac{D(\delta t)^2}{\delta t}.
\end{equation}
We can now go back to the speed steering bound~(\ref{eq:speedBound}) and integrate it on both sides in time from $0$ to $\delta t$ to get
\begin{equation}\label{eq:geometricBound}
\delta t\geq \hbar^2\frac{\langle D(\delta t)^2 \rangle_{B|A}}{\int_0^{\delta t}(\Delta \hat H )^2_{B|A}dt}.
\end{equation}
This bound can be saturated if the system follows the geodesic of the dynamics \cite{abiuso2020geometric}. This, however, cannot be reached by a unitary evolution in the general case. For unitary evolution, we have 
\begin{equation}
    \int_0^{\delta t}(\Delta \hat H )^2_{B|A}dt=(\Delta \hat H )^2_{B|A} \delta t.
\end{equation}
Using this in Eq.~(\ref{eq:geometricBound}), we obtain
\begin{equation}
\delta t\geq \hbar\sqrt{\frac{\langle D(\delta t)^2 \rangle_{B|A}}{(\Delta \hat H )^2_{B|A}}}.
\end{equation}

\textit{Proof of Eq.~(\ref{eq:GHZtime}).}
This state is a mixture given by:
\begin{equation}
\hat\rho = p\ketbra{GHZ^{N+1}}{GHZ^{N+1}} + \frac{(1-p) }{2^{N+1}}\mathds{1},
\end{equation}

where $\ket{GHZ^{N+1}}$ represents the GHZ state:

\begin{equation}
\ket{GHZ^{N+1}} = \frac{1}{\sqrt{2}}\left(\ket{0}_A \otimes \ket{0}_B^{\otimes N}+\ket{1}_A \otimes \ket{1}^{\otimes N}_B\right),
\end{equation}

with one qubit controlled by Alice and the remaining $N$ by Bob. Here, $\ket{0}$ and $\ket{1}$ are the eigenstates of the Pauli operator $\hat\sigma_z$.

Now, consider a local Hamiltonian of the form $\hat J_B^z = \frac{\mu}{2}\sum_i \hat\sigma_z^{(i)}$, where the sum extends over all particles on Bob's side. When Alice measures $\hat\sigma_z$, Bob's particles collapse into a state with minimal energy variance, leading to a quantum speed limit given by:

\begin{equation}
\frac{\Delta^2 \hat  H _{B|A}}{\mu^2}\leq \frac{(1-p)N}{4}.
\end{equation}

On the other hand, when Alice measures in the $\hat \sigma_x$ basis, Bob's state collapses into a mixture of GHZ states that maximizes the quantum Fisher information and, therefore, the mean quadratic velocity of change on the manifold of quantum states:

\begin{equation}
\langle v^2\rangle_{B|A}= \frac{I_{B|A}}{4}=\frac{\mu^2}{\hbar^2} \frac{p^2N^2}{p+2\left(\frac{1-p}{2^N}\right)}.
\end{equation}

Hence, if the visibility parameter $p$  exceeds the critical value $p_c=\frac{2^{N} + \sqrt{2^{N} \left(2^{N} + 32 N\right)} - 4}{8 \cdot 2^{N} N + 2 \cdot 2^{N} - 4}$, Bob will observe an apparent violation of the quantum speed limit, which implies steering from Alice to Bob.

Finally, by employing equation \ref{eq:metroTime}, we can derive a steering condition on the time of evolution, given by:

\begin{equation}
\delta t\geq \frac{\hbar}{\mu} p \sqrt{\frac{N }{\left(2^{1-N}(1-p)+p\right)\left(1-p+N(1-p)p\right)}}.
\end{equation}

This condition is equivalent to the previous one but allows us to use time as a steering witness.

\end{document}